# Decentralized Lifetime Maximizing Tree with Clustering for Data Delivery in Wireless Sensor Networks

Deepali Virmani[1] and Satbir Jain[2]

[1]Department of CSE, BPIT, GGSIPU, Delhi, India, 110085.
deepalivirmani@gmail.com
[2]Department of CSE, NSIT, DU, Delhi, India, 110003.
Jain_satbir@yahoo.com

**Abstract**
A wireless sensor network has a wide application domain which is expanding everyday and they have been deployed pertaining to their application area. An application independent approach is yet to come to terms with the ongoing exploitation of the WSNs. In this paper we propose a decentralized lifetime maximizing tree for application independent data aggregation scheme using the clustering for data delivery in WSNs. The proposed tree will minimize the energy consumption which has been a resisting factor in the smooth working of WSNs as well as minimize the distance between the communicating nodes under the control of a sub-sink which further communicate and transfer data to the sink node.

## 1. Introduction

The current technological advancement has already come to terms with immense potential of Wireless Sensor Network, Which consists of tiny sensor nodes scattered in a region communicating with each other over well defined protocols and transferring information of temperature, humidity etc between each other. To exploit the potential of WSNs various studies have focused on Data aggregation approach which requires data to be collected and processed at a single node prior to its transfer to the parent node. Realization of this data aggregation approach has been a major concern owing to the limited battery life of the sensors which limits the lifetime for which a sensor node remains active. The limited battery leads to disruption in the connections from the network. These disruptions suggest that the design must incorporate topological changes. Since the dense deployment of sensors nodes leads to the detection and transmission of data from the nearby nodes upon receiving a single stimulus. Thus, idea is not to allow the direct transmission of to interested users upon event detection instead aggregating them to remove redundancy. The application domain of WSN is still expanding therefore it is important to support the data aggregation scheme from multiple nodes for simultaneous and fast processing of the data.

In this paper, we focus on the construction of decentralized life time maximizing tree based on clustering. Our scheme consists of three parts, namely, the clustering of nodes by using the Expectation-maximization (EM) [9] algorithm, construction of decentralized life maximizing tree within the cluster [8] , and aggregating the data collected from the WSN nodes by applying a cluster scheduling approach to transfer it , which uses HyMac[1] mechanism.

The rest of the paper is organized as follows. Section 2 describes some related works. Section 3 elucidates our approach. Section 4 has the proposed algorithm. Section 5 we explain our approach with example, section 6 shows the simulation results. Finally, concluding remarks are provided in Section 7.

## 2. Related Works

In recent literature many studies have achieved data aggregation using several approaches namely mobile sink , LEACH(Low-Energy Adaptive Clustering Hierarchy), Directed Diffusion [9].All these schemes have tried to prolong the network lifetime and reduce the energy consumption The mobile sink scheme increases the network lifetime to four times as compared to the network in which sink is static but it suffers from serious shortcomings , it leads to an increased physical delay owing to the slow physical mobility of the sink then the wireless communication. It exhausts the battery life unnecessarily. LEACH is a self-organizing, adaptive clustering protocol. To have minimum energy consumption, nodes in LEACH are grouped into a number of clusters based on their battery usage. Each cluster has a cluster head, which communicates with every node of that cluster. The sink aggregates data, transmitted by cluster heads, from other nodes. Since a cluster head loses energy due to repeated transmissions, the cluster head is re-selected based on the residual energy, as a consequence it prolongs the network lifetime. Directed Diffusion involves two types of messages, namely, the "Interest" message and the actual data messages. To aggregate data by using Directed Diffusion, the sink node broadcasts an "Interest" message that consists of a time-to-live value, and also the addresses of the source and destination nodes. The destination node on receiving the request transmits appropriate data message to the source having the sensed data. If the downstream nodes cannot be reached by the "interest" message from the current source then the current destination becomes the source node by changes its address, reduces the time–to-live value and rebroadcasts the "Interest "message.

## 3. Proposed Approach

We have discussed before that our main focus has been on distance minimization between the nodes, minimization of energy utilization and efficient utilization of bandwidth. Considering the problem sensor network is divided into clusters using an EM [9] algorithm based on the close proximity of the nodes and a decentralized life maximizing [8] tree is constructed within the cluster choosing a parent closest to the sink node to serve as sub-sink. Once the sub sink is chosen, scheduling of cluster is done using FDMA approach. That is assigning a range of frequencies from the available one to the sub-sink. The frequency ranges can be re-allotted to new clusters from the free pool or assigning a half frequency to a sub-sink from a low data rate transferring cluster. The sink will broadcast a topology packet containing information of the network as which source nodes are attached to which sub-sink node as per their location [4]. By making use of the hybrid TDMA/FDMA channel access technique [1], the sink node broadcasts a schedule packet informing others about their time slots as well as their channel frequencies for exchanging messages. But, our new idea lies in single sleep awake concept in which the source nodes wake up only once to listen and to transmit and rest of the time, they will remain in sleep state. We incorporate a concept of LPL (low power listening),the nodes are in LPL [2] state all the time to gauge topology changes and if there is a topology packet coming their way they wake up and make necessary changes. The tree construction follows a decentralized method [8]. To overcome the problems in [1]and to decrease the interference respectively we propose this new method. We will assign specific frequency slots based on attributes of the sensor nodes [6], with fixed interference ranges so that they can send their data in scheduled time in slotted frequency. Once the sender finishes sending, same frequency can be assigned to some other source accounting for the increase slots needed and also minimizing interferences.

To preserve the functional lifetime of all sources and efficient utilization of the energy of the source nodes, a Decentralized life maximizing tree construction algorithm [8] was studied, the DLMT [8] constructs a tree by selecting highest residual energy parent node to act as a center of data aggregation. The DLMT [8] construction algorithm arranges all nodes in a way that each parent will have the maximal-available energy resources to receive data from all of its children. Such arrangement extends the time to refresh the tree and lowers the amount of data lost due to a broken tree link before the tree reconstructions. The DLMT [8] algorithm can be further improved by considering distance also as a factor. In the proposed method we also include distance between the sensor nodes. Transmission distance has a major impact on the working of sensor network because the required power of wireless transmission is proportional to the square of the transmission distance. We follow the approach of clustering of nodes based on EM [9] algorithm. The EM algorithm includes minimizing the sum of the squares of the distances between nodes and cluster centroids. Therefore, we use the EM [9] algorithm to group the WSN nodes into K clusters on the basis of distance. We apply the concept of EM [9] algorithm initially and then use a new form of decentralized life maximizing tree, DLMT [8] algorithm accordingly. The cluster formed using the EMD algorithm goes through our proposed algorithm called Decentralized Lifetime Maximizing Tree using Clustering based energy and distance (DMLTC), which creates trees within the clusters already created .The choice of the tree is based on the minimum distance of the sub-sink from the sink.

The tree that we get after application of both the algorithms is efficient in terms of distance as well as energy, now to improve the bandwidth utilization we apply a method of FDMA_SINK ( ) that allots range of frequencies to the sub-sink for data transfer and also checks for the efficient utilization of bandwidth by assigning half frequency range from a low data transfer cluster to a new one.

Finally, after frequency allotment we implement HyMac [10] algorithm which provides fixed time slots to nodes to transmit sensed data, the sub-sink remains in a LPL state and listens for receiving data from the children, the child nodes awake once and start the synchronous data transfer to the sub-sink which further sends sensed data to the sink in the same assigned time slot.

## 4. PROPOSED ALGORITHM

### 4.1 EM Algorithm

The following algorithm that we use divides the network into K clusters. It has been renamed from EM [9] to EMD (Expectation Maximization on Distance) Algorithm.

K : The number of clusters

πk: The mixing coefficients of the kth cluster

μk: The 2-dimensional vector indicates the mean of the kth cluster

Σk: The 2 × 2 covariance matrix of the kth cluster

---

Algorithm 1: EMD Algorithm

The mobile sink node groups all nodes into K clusters by using the EMD algorithm in the following manner.

1: Initialize μ, Σ, π and the convergence criterion θEM, and evaluate the initial value of P:

$$P = \sum_{n=1}^{N} ln \left\{ \sum_{k=1}^{K} \pi_k N(x_n | \mu_k, \Sigma_k) \right\}$$

Where N is the number of nodes.

---

The above presented algorithm groups the sensor nodes into clusters based on their distance and hence ensure that the nodes in a cluster have a close proximity which will lead to minimization of data transfer delays. The algorithm can be iteratively used to account for any new nodes coming up in the network.

## 4.2 Decentralized Life Maximizing Tree with Clusters: Algorithm

After the nodes have been grouped into clusters using EMD algorithm, our next step is to apply DLMT [8] in a new form, DLMTC which further takes into consideration both the energy and the distance factors. It constructs trees within the clusters considering energy factor but the final choice of the tree is done by taking into account the distance of the sub-sink or parent from the sink node. The algorithm can be applied iteratively to compensate for the broken links or when a new node is added to the existing network. The presented tree construction leads to minimum delays in data transfer as well as judicious use of the available energy of the nodes.

## 4.3 Proposed DLMTC Algorithm:

This algorithm compares the previous tree constructed based on their distance from the sink and chooses the suitable one with minimum distance from the sink.

---

BestDLMTC (DLMTCi, DLMTCj, di, dj)

1. if rows in ( DLMTCj ) > rows in ( DLMTCi )

2. return true

3. if [ rows in (DLMTCj ) = rows in ( DLMTCi ) ] and [ DMLTEj > DLMTC_Ei ]

4. return true

5. if [ rows in ( DLMTCj ) = rows in ( DLMTCi ) ]and [ DMTLEj = DLMTC_Ei ] and [ tree depth of DLMTCj < tree depth of DLMTCi }

6. return true

7. if [ rows in ( DLMTCj ) = rows in ( DLMTCi ) ] and [ DMLTEj = DLMTC_Ei ] and [ tree depth of DLMTCj = tree depth of [ DLMTCi ] and [ ej > ei ] and [dj < di]

> 8. return true
>
> 9. if [ rows in( DLMTCj ) = rows in( DLMTCi ) ] and [ DMLTEj = DLMTC_Ei ] and [tree depth of DLMTCj = tree depth of DLMTCi ] and [ej = ei] and j < i
>
> 10. return true  Else if return false.

## 4.4 Applying FDMA_SINK on the DLMTC Tree

Scheduling using FDMA System Model:

All the frequencies available in the band are divided in frequency ranges(R) based on the number of clusters (K), initially, all sub-sinks send a synchronous message to sink requesting allotment of the frequencies, sink then checks for the availability of the frequency ranges. If it is available it is assigned. Otherwise sink may withdraw a frequency from some other cluster if its data transmission rate is low and assign that to some cluster. We propose an algorithm FDMA_SINK () with parameters (K, number of clusters, Fi, frequency band, i ,Cluster),SSi is subsink of ith cluster.

> Algorithm 3: FDMA_SINK (DLMTC Treei,Ki,Fi,Ci)
>
> 1. Calculate the range $R_i = F_i/K_i$
>
> 2. For each S $S_i \in C_i$ where i ← 1 to k
>
> 3. Assign $R_i$
>
> 4. If SSi receives the last packet then
>
>    Withdraw the frequency and add it to the free pool.
>
> 5. End for
>
> 6. Request for new Fi allotment
>
> 7. If available from free pool, Assign from free pool
>
>    Else withdraws half the frequency range from a low data rate transfer cluster and assign to the subsink.

## 4.5 Cluster Scheduling Using HyMac

Scheduling algorithm is applied on the DLMTC tree having the base node as its root. As each node Ni is traversed by DLMTC, it is assigned a default time slot and a frequency using FDMA_SINK () function discussed before. Then the possibility of having an interference with any of its same-height previously-visited one-hop AND two-hop neighbours is checked. If a conflicting neighbour Nj is found for Ni, the algorithm checks whether Ni and Nj are siblings. If so, Ni will be assigned a different time slot than that of Nj. If they are not siblings then Ni will be assigned a different frequency than that of Nj , allowing both Ni and Nj to send messages to their parents at the same time slot but in different channels. When DLMTC is about to start a new level (height) of nodes the default time slot number will be increased by one. Once all nodes are processed according to the above heuristic, the entire time slot assignments will be inverted such that the slot number assigned to every node is smaller than that of its parent. This inversion is done as following:

$t_{new} = t_{max} - t_{current} + 1$

(1)

Algorithm 4: HyMAC with clusters Scheduling Algorithm

```
Require: A Graph of Sensor Network Topology
Ensure: An scheduled Tree of the Given Network
1: ENQUEUE (Q, S)
2: while Q is not empty do
3:      v ← DEQUEUE (Q)
4:      timeSlot[v] ← currentTimeSlot
5:      FDMA_SINK () /* assigning channels */
6:      for all Visited same-height 1-2-hop nbr n of v do
7:          if parent[n] == parent[v] or #Channel >= available chnls then
8:              if timeSlot[v] = timeSlot[n] then
9:                  timeSlot[v] ← timeSlot[n] + 1
10:             end if
11:         else
12:             if timeSlot[v] = timeslot[n] and channel[v]=channel[n] then
13:                 channel[v] ← channel[n] + 1
14:             end if
15:         end if
16:     end for
17:     for all unexplored edge e of v do
18:         let w be the other unvisited endpoint of edge e
19:         parent[w] ← v
20:         height[v] ← height[w] + 1
21:     end for
22: end while
```

Where t new is the new inverted assigned slot, tcurrent is the current slot number assigned to the node and tmax is the total number of assigned slots. Note that such an assignment allows the data packets to be aggregated and propagated in a cascading manner to the base station in a single TDMA cycle. The complete steps of the overall process are presented in algorithm 4.

FINAL ALGORITHM:

Require: Set of sensor nodes

Step 1: Apply EMD () /* Creates clusters */

Step 2: Creating decentralised Maxlife tree using DLMTC ()

Step 3: Choosing the life maximising tree using Best DMLTC ()

Step 4: Scheduling Using HYMAC () which calls FDMA_SINK for scheduling based on time slots and frequency range.

## 5. An Illustration of our approach:

After the cluster formation using EMD algorithm, we apply DLMTC [9] on the given cluster to create decentralized MaxLife tree (DMLTree) based on their energy. After applying this algorithm we get several trees with different parents, choice of which depends on the energy of the node and distance of the node from the sink. When a node with maximum energy and minimum distance is found the node is taken as SubSink node where data aggregation is then performed. The concept is shown by assuming the trees formation in figure 1-4.

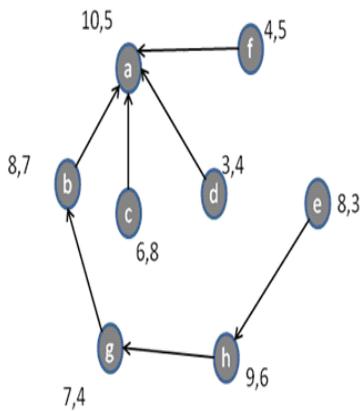

Fig.1 Tree construction within cluster C1 with a as parent

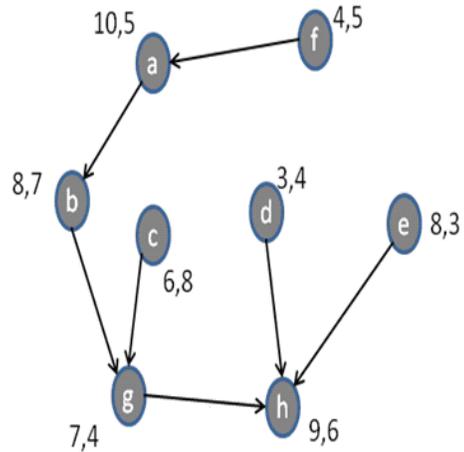

Fig. 3 Tree construction within cluster C1 with g as parent

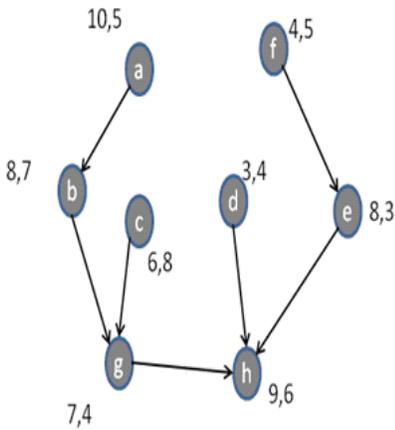

Fig.2 Tree construction within cluster C1 with h as parent

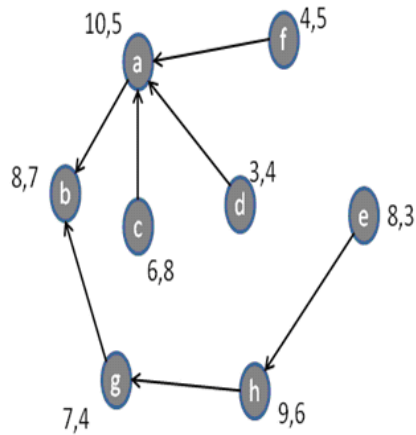

Fig. 4 Tree construction within cluster C1 with b as parent

The best tree found here is the one with node 'a' as the parent (fig.1); hence the node is taken as a SubSink for the cluster C1.

## 6. Simulations Parameters

We implemented our tree construction modules on top of Forwarded Diffusion in the J-Sim network simulator (the J-Sim comes with diffusion support). In all of our experiments, a square sensor field with each side measuring X meters is being considered. A number of N identical nodes, ranging from 50 to 300 in the increment of 50, are randomly deployed in this sensor field such that the average node density is kept at $\lambda = 55/1652$ nodes per meter square, a parameter which we borrowed from Forwarded Diffusion [10, 11]. Furthermore, there are five sinks randomly deployed in the field and sources are randomly chosen among the nodes, subject to the conditions that SR=10% of N and the sources have to be interconnected to each other (to model a single stimulus). Each node is assumed to have a radio range of 45 meters. We considered an event-driven data sensor network throughout all our experiments. To model the periodic transmissions, each source generates random data reports of size fixed at 138 bytes in constant intervals of DR = 1 packet/second. To introduce some randomness, data start to be generated only after a time randomly chosen between t = 0 to 5 seconds. The data are collected at the root, if they exist, and are sent to the sinks. We assign each source with an initial energy that is randomly chosen between 10 to 18 Joules in order to keep the total simulation time at a reasonable limit. In all of our experiments, all other nodes are given an initial energy that is greater than that of any event source such that their absence in the network, due to energy depletion, does not affect the functionalities of any participating sources during data collection. Lastly, the idle time power, receive time power and transmit power dissipation are set at 40, 400 and 680 mW respectively. We assume a negligible energy cost to process and aggregate incoming data reports. To trace the energy, an application that logs the residual energy of each node in constant intervals of 550 ms is employed. The J-Sim simulator implements a 1.6 Mbps 802.11 MAC layer. Since Forwarded Diffusion is chosen as our routing platform,

6.1 Average Dissipated Energy (ADE): ADE measures the average amount of energy consumed throughout the entire simulation. This metric computes the average work done in delivering periodic data to the sink/ Root over a simulation run. As shown in fig. 5 energy conservation is more with DLMTC as compared to DLMT that is with clusters we are able to preserve more energy.

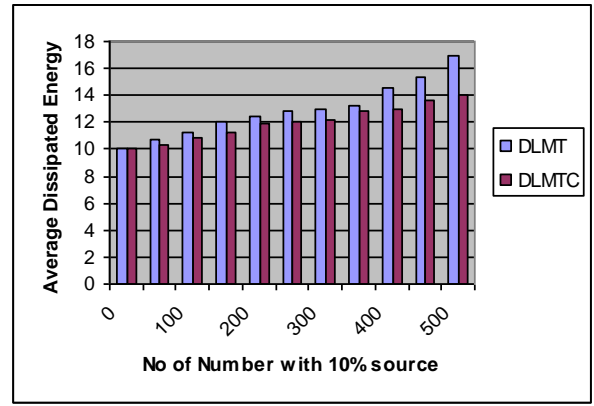

Fig. 5 Average dissipated energy

### 6.2 Average Network Lifetime (ANLT):

In order to study the impact of DLMT, CLMT and E-Span on the lifetime-savings, we measure the node lifetime of each source as a function of network size for DLMT, CLMT and E-Span respectively. Each node is assigned number of with an initial energy that is randomly chosen between 12 to 18 Joules so as to limit the total simulation time at a controllable range. Reference [13] shows that DLMT enhances the network lifetime in comparisons with other trees constructed for the same purpose. But now these simulation results shown in fig. 6 we are able to prove that including clusters in the DLMT makes the node alive for enhanced time.

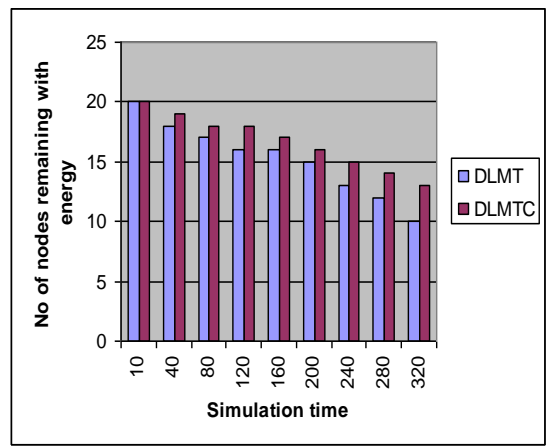

Fig.6 Average network lifetime

### 6.3 Average Delay

Average delay measures the delay between transmitting data from each source to each of the sinks. This is the basic limitation of DLMTC average delay is maximized as extra time is required in setting cluster head and other parameters. Once the clusters are formed the process speeds up and energy is preserved. Fig. 7 shows the results that at start of simulation delay is maximum as the process of formation of clusters is going on , once the cluster formation is done then the process speeds up minimizing the delay.

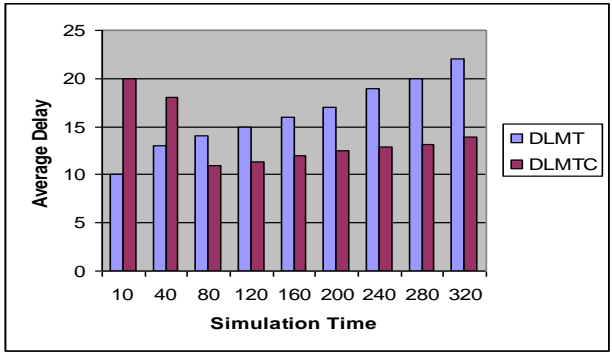

Fig. 7 Average delay

### 6.4 Bandwidth Utilization

As bandwidth allocation with DLMTC is based on allotment of frequency from the free pool of frequencies or by withdrawing half frequency from cluster with low data transfer rate. Results as shown in fig. 8 prove that bandwidth utilization is almost double for DLMTC as compared to DLMT.

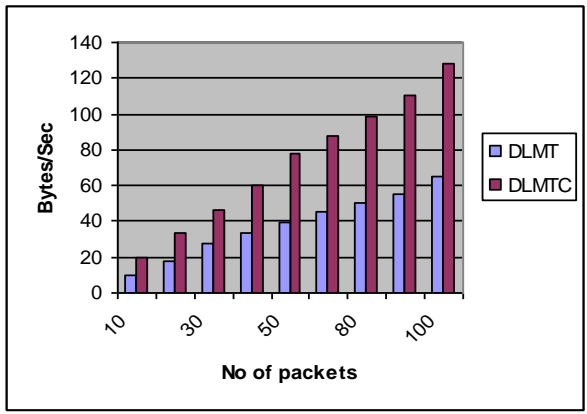

Fig. 8 Bandwidth Utilization

## 7. Conclusions

In this paper we proposed the Decentralized Lifetime Maximizing Tree with clustering construction algorithm. Clustering on the basis of distance ensures close proximity of the nodes and thus leads to reduction in data transfer delays. Energy conservation by waking nodes once instead of twice leads [5] to further reduction in data transfer delays, thus utilizing the energy available effectively. Efficient utilization of bandwidth achieved through allotment of frequencies from the free pool or withdrawing half frequency from cluster with low data transfer and assigning it some other. Efficient utilization of energy is achieved by using HyMAC technique. Simulation results prove enhancement if network lifetime, reduction in energy consumption and minimization of average delay.